\documentclass[runningheads,a4paper]{llncs}
\usepackage{graphicx}
\usepackage{amsmath}
\usepackage{amssymb}
\usepackage[hidelinks]{hyperref}
\usepackage{algorithm}
\usepackage[noend]{algpseudocode}
\usepackage{bbding}
\usepackage{breqn}
\usepackage{tikz}
\usepackage{pgfplots}
\usepackage{blindtext}
\usepackage{dirtytalk}
\usepackage{multirow}
\usepackage{url}
\usepackage{enumitem}
\usepackage{float}
\usepackage{caption}

\newcommand{\keywords}[1]{\par\addvspace\baselineskip
\noindent\keywordname\enspace\ignorespaces#1}

\begin{document}

\mainmatter  

\title{Automation of Android Applications Testing Using Machine Learning Activities Classification}

\titlerunning{Automation of Android Applications Testing with Machine Learning}

%
%
\author{Ariel Rosenfeld$^{1,\dagger}$\and Odaya Kardashov$^{2}$\and Orel Zang$^{2}$}
\authorrunning{Rosenfeld, Kardashov and Zang.}

\institute{$^{1}$ Dept. of Computer Science and Applied Mathematics,\\ Weizmann Institute of Science, Rehovot, Israel\\
$^{2}$ Dept. of Computer Science, Bar-Ilan University, Ramat-Gan, Israel\\
$^\dagger$ Corresponding author, arielros1@gmail.com
}

%
%

\toctitle{Automation of Android Applications Testing with Machine Learning}
\tocauthor{Rosenfeld et al.}
\maketitle
\begin{abstract}
Mobile applications are being used every day by more than half of the world's population to perform a great variety of tasks. With the increasingly widespread usage of these applications, the need arises for efficient techniques to test them. Many frameworks allow automating the process of application testing, however existing frameworks mainly rely on the application developer for providing testing scripts for each developed application, thus preventing reuse of these tests for similar applications. In this paper, we present a novel approach for the automation of testing Android applications by leveraging machine learning techniques and reusing popular test scenarios. We discuss and demonstrate the potential benefits of our approach in an empirical study where we show that our developed testing tool, based on the proposed approach, outperforms standard methods in realistic settings. 

\end{abstract}

\section{Introduction}
Mobile devices become a key component in our lives, with more than half of the world's population now owning one \cite{2017Digital}. More than five million applications have been developed so far \cite{NumberOfApps}, making them the main productivity feature of these devices. These applications have completely changed the way we handle everyday activities, communicate with each other and perform many tasks \cite{akour2016mobile}. As mobile devices become more popular, thus arise the need for efficient techniques for testing their applications. The large fragmentation of the Android market, as well as the diverse set of scenarios in which a mobile application can be used,  make testing new applications an expensive, time-consuming and complex process \cite{huang2012remote,akour2016mobile}. Unfortunately, mobile applications are proving to be bugs-prone mainly due to developers' unfamiliarity with mobile platforms \cite{hu2011automating}. A study by Maji et al. \cite{maji2010characterizing} has found that mobile applications tend to significant amount of defects and bugs. 

Traditionally, companies used \textit{manual} testing methods for testing their applications. However, recently, a growing number of companies and organizations use automated tools for their testings  \cite{gao2014mobile}. A recent survey conducted among 644 stakeholders that have direct influence on the test automation procedures in their organizations \cite{FrameworksSurvey}, shows that these companies can use up to $4$ automated tools in parallel to maximize the likelihood of detecting bugs. Test automation became the standard, with many solutions and frameworks that allow automating the process of application testing. These frameworks enable a developer to write code for application functional testing and executing this code in test specific scenarios. The problem with this approach is that these tests are hand-coded for specific applications and specific scenarios, and each new application requires spending many resources to reuse these tests. In addition, these tests require high maintenance since every change in the application should be reflected in the pre-coded tests.

In this paper, we present a novel approach for automatic testing of Android applications in order to find as many functional bugs as possible. Our approach is based on the premise that different activities in an Android application share a similar interface structure. In order to use this similarity for our benefit, we use machine learning techniques to classify each activity in the application into one of seven pre-defined activity types which are identified in this work. For each classified activity, we then run specialized tests, at user interface level, that were coded to utilize the fact that we know the activity's structure and desired behavior. We have implemented this approach and developed an add-on in Java for the TestProject\footnote{\url{http://testproject.io}} test automation framework that uses the Appium\footnote{\url{http://appium.io/}} open-source framework as a bridge between the mobile device and our code. TestProject allows a developer to build, deploy and execute automated testing by utilizing popular open-source frameworks for both Web and Mobile applications. The platform includes hundreds of add-ons for automated testing which are freely available. Our developed add-on, named ACAT, standing for \say{Activities Classification for Application Testing}, will be available to install via TestProject add-ons store.

To evaluate our approach, we conducted an experiment in which we executed our add-on on different applications. We found that the ACAT add-on shows great ability in exploring the application and testing its key components without prior knowledge about the application. This lets the developer focus on the development of the application and not on writing standard tests. The use of machine learning for testing applications is, to the best of our knowledge, a novel approach which has yet to be fully explored. 

The rest of this paper is organized as follows. Section \ref{sec:relatedwork} discusses related works. Section \ref{sec:approach} presents our approach. Section \ref{sec:evaluation} describes our experiment and its results and Section \ref{sec:discussion} provides a discussion about the results. Finally, Section \ref{sec:conclusions} provides conclusions and future work.

\section{Related Work}
\label{sec:relatedwork}
Producing a tool that will allow testing any arbitrary mobile application automatically is an extremely challenging problem, perhaps nearly as difficult as the underlying general software testing automation task. Throughout the years, there has been an extensive study in the field of testing automation of \textit{desktop applications}. However, a recent study by Hu et al. \cite{hu2011automating} has discovered that many of the mobile application bugs are unique and tend to be different from the ones presented in traditional desktop applications, mainly due to the inherent difference in architecture and development methodologies. Thus, traditional approaches for desktop applications testing cannot be naturally translated into mobile applications.

The need for efficient mobile application testing methods has yielded many testing automation frameworks, including tools like Appium\footnote{\url{http://appium.io/}}, Selendroid\footnote{\url{http://selendroid.io/}} and Robotium\footnote{\url{https://github.com/RobotiumTech/robotium}}, to name a few (a recent survey is available at \cite{gao2014mobile}). These frameworks allow a developer to write testing scripts in her programming language of choice, and later run these scripts over and over again to check the application in different user behavior scenarios. The main limitation of such tools is that these manually constructed scripts are coded for a designated application in mind, therefore the developer is bound to invest ample time in reusing these tests for new applications and to accommodate for changes in the functionality of existing applications.

In order to reduce the need for writing redundant testing scripts for mobile applications, which share many common characteristics, a great amount of research efforts have lately focused on the development of automated testing techniques and algorithms to allow for testing applications automatically. In this paper, as with most of recent papers in the field, we focus on the Android platform  \cite{choudhary2015automated}. The choice to use the Android platform  is mainly due to the fact that it is the most common mobile operation system on the market to date and due to its open-source nature that allows the academic community to get full access to the applications and the platform source code. Moreover, the large variety of Android models and versions on the market make the test automation task significantly important.

Testing an Android application using automated testing techniques is commonly executed by running the designated application while generating user interface events, which simulate a user behavior including actions like clicks, scrolls and swipes. The challenge of this task is to generate as much \textit{relevant inputs} as possible in order to explore the application with maximum coverage. According to a recent study by Choudhary et al. \cite{choudhary2015automated}, which has conducted a comprehensive overview of the main existing Android applications testing tools that have been proposed and developed in academic papers, we can categorize each of these tools into one of three approaches:
\begin{itemize} \item \textbf{Random exploration approach:} Tools which use this approach generate user interface events in a \textit{random fashion}, executing them one by one on the application user interface. The main use of this approach is to test the application robustness, as most of these events are ones that the average user is not likely to perform. Random test input generators are easy to use and are particularly suitable for so-called \say{stress testing}. On the downside of this approach, random tests are prone to get stuck with repetitive events and are not likely to get a good coverage of the application, due to their random nature. In this category we can find tools like Monkey \cite{Monkey}, Dynodroid \cite{machiry2013dynodroid} and DroidFuzzer \cite{ye2013droidfuzzer}. \item \textbf{Model-based exploration approach:} Tools which use this approach build a model of the application's GUI in order to utilize it for building a sequence of user interface events that maximize the exploration coverage. The model is usually a finite state machine which its states are the application different screens and its transitions are the different possible user interface events. These tests trigger all the different possible user interface events (i.e., the finite state machine's transactions) on all the different screens (i.e., the finite state machine's states) and ends when all the user interface events that can be triggered are leading to already visited screens. The advantage of this approach is that it tends to get a good coverage of the application as triggered user behaviors are unique. The limitation of this approach is that only changes in the application GUI are reflected as new states in the model. However, many times user interface events change the internal state of the application, thus making these models miss and avoid certain exploration routes. In this category we can find tools like GUIRipper \cite{amalfitano2012using}, A$^3$E-Depth-First \cite{azim2013targeted} and Swifthand \cite{choi2013guided}, which uses machine learning techniques to learn a model of the application's GUI and guide the generation of user input sequences based on this model. \item \textbf{Systematic exploration approach:} These tools generate unique user behaviors by dynamically analyzing the application's \textit{source code}. The strength of this approach is that it can leverage the source code to generate tests to reveal previously uncovered application behavior. The downside of the approach is significant scalability concerns. In this category we can find tools like Sapienz \cite{mao2016sapienz}, EvoDroid \cite{mahmood2014evodroid}, A$^3$ E-Targeted \cite{azim2013targeted} and ACTEve \cite{anand2012automated}.
\end{itemize} 

All of the above approaches have been successfully deployed in different experiments to have shown to produce a good coverage of the applications' state-space. Nevertheless, they all share a common prominent limitation: these approaches aim to find only \textit{technical} bugs and defects in the application, meaning real-time application crashes which are caused by uncaught exceptions thrown in the code \cite{choudhary2015automated}. However, many of the bugs presented in today's applications are related to the program logic (e.g., a login screen that can be by-passed without entering valid username and password, an email-composing screen that allows sending emails to an invalid email address, etc). Hu et al. \cite{hu2011automating} present an empirical study of common Android bugs. The authors find that logical bugs are about $10$ times more prevalent than technical bugs.

Another significant limitation of the above three approaches is that, in general, they will not be able to reach screens of the application that need a specific input in order to advance. Specifically, the above approaches will probably get stuck when reaching a login screen, being unable to cover a (potentially) significant part of the application. Choudhary et al. \cite{choudhary2015automated} points out this problem as a future research direction, saying that allowing tools to explore the application in presence of login forms and similar complex inputs, which may be hard to generate randomly or by means of systematic techniques, will help explore new  behaviors. In this work, we suggest to overcome this limitation by using machine learning techniques that enable our new approach to test new, previously unexplored, logical conditions in the screen using our pre-defined set of expected behaviors, as we will describe later on. Moreover, using very limited inputs supplied by the programmer, previously impassable screens can now be passed by feeding these inputs at the right time. Thus, using our approach, one can test applications more comprehensively, finding new logical bugs and reaching new states in the application.

\section{Approach}
\label{sec:approach}
\subsection{Motivation for Activities Classification}

The key element of our approach stems from the difference between testing desktop and mobile applications. While desktop applications come in an endless amount of shapes and forms, the structural scope of mobile applications is naturally more limited \cite{SOFIA}. An Android application is, at its core, a series of different screens which are connected using user interface buttons. The official Android development guide defines each of these screens as an \say{Activity}\footnote{\url{https://developer.android.com/reference/android/app/Activity.html}}, which is a single window in the application. An Android activity is a group of different user interface elements from the Android development kit which are organized in a hierarchic structure. While these elements vary in their specific purposes, we can categorize them into two main groups: 1) Elements which are directly visible to the user on the screen and allow him to interact with them by hand gestures, such as clickable buttons, lists of items which can be scrolled up and down and text fields; 2) Elements that are not directly visible to the user on the screen, but rather control the layout of other user interface elements in the activity, such as arranging them horizontally in a single column or vertically in a single row.

In a session titled \say{Structure in Android App Design} given at the Google I/O 2013 developer conference \cite{AndroidAppStructure}, Nagel and Fulcher discuss common patterns in designing activities for Android Applications. They introduce various structures of elements arrangement, explaining that using common and more simple activities structures will help making the application more predictable and understandable to the user and thus, more pleasing to use. The fact that many different Android activities share the same structure suggests that these activities may require similar treatments as for their testing. Therefore, by classifying an activity into a certain class of activities, we can derive which tests should be performed automatically. The problem can be naturally translated into a Multiclass classification problem \cite{aly2005survey}, which is a common branch of machine learning. Namely, given an instance of an activity, we seek to classify it into one of seven pre-defined classes.

\subsection{A Study of Activities Types}
\label{sec:activitiesStudy}

Narrowing down all of the possible activities into a finite list of types is an open question for future study, as it would require a more comprehensive study. In the scope of this work, we performed a preliminary study of 100 Android applications from the Google Play store, by manually searching for common patterns, structures and behaviors in the different activities. Based on our preliminary study, we identified $7$ activity types which can be divided into two  groups: 1) Activity types which have been the most common ones among the 100 studied applications. 2) Activity types which have a notable structure and a naturally-anticipated functionality.

Activity types in the first group:
\begin{itemize}
\item \textbf{Splash Activity:} Splash activity is the screen displayed when opening the activity, which usually displays an image or text while the application is loading in the background. Most of the applications that we examined had a splash activity. When a splash activity exists, it is always the first screen in the application. Nevertheless, there is still a need to classify the first activity in the application. For many lightweight applications, this screen may be displayed for only a fraction of a second, which may result in incorrectly classifying the second screen of the application as a splash screen if the classification is done in a na\"ive way. The test for this activity will be to make sure that the application can advance from this screen to the next one.
\item \textbf{Advertisement Activity:} The vast majority of the Android applications are free to download due to many reasons listed in \cite{AppPricing}. Instead, for making profits, the Android developers commonly incorporate advertisements in their applications. These advertisements can pop up anywhere and at anytime in the application, which makes the classification challenging. Identifying these activities is important as clicking on these advertisements during testing will likely exit the designated application. Thus, after classifying an activity as an advertisement activity we need to carefully close it and make sure that we stay within the scope of the application.
\item \textbf{Login Activity:} Many modern applications require a valid username and password in order to use most of the application's services. Therefore, these applications contain a screen which allows the user to enter its username and password to connect to their server(s). We designed a test for this activity that verifies that the user cannot bypass the login screen by leaving the text fields empty, or by entering incorrect credentials (e.g., using random strings). After that, the test verifies that the login screen can be passed by using valid details (the test should be provided with valid username(s) and password(s)).
\item \textbf{Portal Activity:} Many of today's communications media, such as websites, newspapers and TV channels, have a designated mobile application that allows the user to access the media content on her phone. The portal activity is the \say{hub} screen of these applications, and thus we designed a test that verifies that the screen can be swiped left and right in order to reach different sections of the portal and that an article can be opened from this screen.
\end{itemize} 
Activity types in the second group:
\begin{itemize}
\item \textbf{Mail Activity:} Mail applications, which are very common on mobile devices \cite{BestMail},  have a well defined purpose with a limited number of possible actions. This allows us to design global tests for every mail application. The mail activity is the \say{hub} screen of these applications, with functionalities as managing the inbox mails and sending new mails. We designed a test for this activity that browses through the inbox mails, tries to open a mail from the list (at random) and scrolling through the mails content.
\item \textbf{Browser Activity:} Web browsers are one of the most important applications, as they allow the user to access websites on their mobile device \cite{BestBrowser}. These applications share a very specific functionality which can be translated into a uniform test which verifies that the user can reach several websites through the activity, use the back, forward and home buttons and opening a new tab.
\item \textbf{To Do List Activity:} 
To-do list applications usually share a common purpose which is to keep track of the user personal list of tasks. Thus, the test for this activity verifies that the user can add new tasks to the list and check them as done.
\end{itemize} 

As we continue to describe our approach, one must consider the major challenges in designing activity tests. These tests cannot be hard-coded for a specific application in mind as they should fit  different activities which share the same type in as many as possible different applications. Different programmers develop different applications, each have her own way and style of designing the application. For example, the developer may refer to different elements on the screen by using  varying  \textit{resourceID}s. As a result, when manually coding a test for an Android application, referring to the correct resourceIDs is not a problem. Unfortunately, this is infeasible in our case as we propose tests for general activities which we do not know the resourceIDs for their elements in advance. In order to overcome this problem, we built lists of associative words for different elements we expect to find in a test. When searching for a specific element on the screen, e.g., a close button for an advertisement, we iterate over the clickable elements on the screen. For each clickable element we check if its resourceID contain at least one word from the list: [\say{close},\say{discard},\say{shut},\say{hide},\say{no}]. If so, we assume that this button is the close button of the advertisement. From our experiments (see Section \ref{sec:evaluation}), this method is proving to be very efficient, as the resourceIDs developers give their elements tend to be very predictable. The rational for the above is that developers themselves want to give informative names to their elements, as this will help simplify the code maintenance.

In this paper we have decided to focus on 7 common activity screens as we identified in our preliminary study (see Section \ref{sec:activitiesStudy}). However, note that our approach can be readily amended with additional activities. This is left for future work.

\subsection{Building the Features Vector}
\label{sec:features}
Each activity can be characterized by a large number of features, which are all related to the user interface elements it contains such as the different classes of the elements, their set of attributes,  their relative location in the activity, the number of elements presented in the activity, etc. Additionally, an activity can be characterized if it contains a navigation drawer, which is a panel that displays the application’s main navigation options on the left edge of the screen. It is hidden most of the time, but it is revealed when the user swipes a finger from the left edge of the screen or, by clicking on a designated button.

\begin{figure}[H]
  \includegraphics[width=\linewidth]{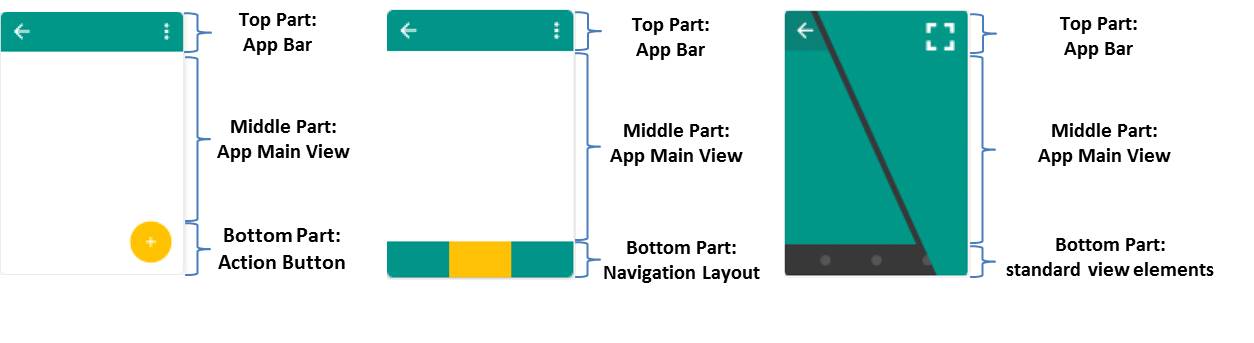}
  \caption{Screen divisions across common activities templates}
  \label{fig:screenDivisions}
\end{figure}

While constructing the features vector, we had to decide which elements are the most informative and may differ between different types of activities. In our preliminary study, as described in Section~\ref{sec:activitiesStudy}, we  noticed that an activity can be identified mostly by its visible elements, namely, the elements which the user can interact with directly. This correlates to the fact that the official Android development guide specify that almost all activities interact with the user. As a result, user interactive elements are assumed to adequately represent the activity. Furthermore, by examining the basic activity templates from the Android studio activity design guideline\footnote{\url{ https://developer.android.com/studio/projects/templates.html}}, we noticed that each activity screen can be artificially divided into 3 parts: the top, the middle and the bottom. We use the following heuristic division of the screen: 20\%-60\%-20\% from top to bottom, as depicted in Figure \ref{fig:screenDivisions}.

Therefore, we focus on the following interactive elements groups which can appear in each of the three activity's parts: 
\begin{itemize} 
\item Clickable elements: Elements that are responsive to the user touch click.
\item Horizontal swipeable elements: Elements that can be swiped by the user left and right. 
\item Vertical swipeable elements: Elements that can be swiped by the user up and down. 
\item Text field elements: Elements that the user can type text into them.
\end{itemize}

We use the number of elements from each of the above element groups in each of the three activity parts as the first set of features.  Namely, the first group of features contains $12$ features, where each represents the number of the elements of each of the $4$ element categories presented above in each of the 3 parts of the screen. 

The second group of features contains 2 features. The first one is the number of general elements on the screen, no matter what class they are or where they are located. The second is the number of  'long-clickable' elements on the screen, meaning elements that respond to the user holding them for a couple of seconds, such as an image element that holding it for a couple of seconds will mark the image and display options such as saving or sharing it. Based on our experience in applications development and our activities study, these features are less prominent and thus we do not divide them into different parts of the screen.

The final group contains one feature, which is a boolean variable  set to true if the activity contains a navigation drawer. We can determine if an activity contains a drawer by checking if it contains an element of class DrawerLayout, which is the default navigation drawer the Android development kit provides. However, this feature is still hard to derive as some applications implement a different drawer than the default one, and as a consequence they do not contain a DrawerLayout element. To overcome this problem, when we are scanning for clickable elements on the screen, we check the resourceID of each one of them. If the resourceID contains one word from a pre-defined constant list of words that indicates a drawer button, such as \say{drawer}, \say{menu}, \say{sidebar} and so on, we assume that this button opens a drawer menu in the activity and thus the activity contains a drawer.

Overall, we use 15 features as described above.

\subsection{Constructing a Dataset} 
In order to construct a dataset to train and test our classifier, as discussed in Section~\ref{sec:classifier}, we needed to obtain a large set of Android applications' activities and perform feature extraction and labeling. Extracting the features of an activity is a hard task, due to the fact that many of the elements in an activity are invisible and cannot be identified just by observing the activity display on the device. To overcome this problem, we use the \say{TestProject Elements Spy} tool, which allows developers to scan and inspect the user interface elements of an Android activity. We have implemented an automated script which extracts the features of a given Android application, as defined in Section~\ref{sec:features}, using the Elements Spy tool  and saves them to the dataset file in a textual format. Although the extraction of the features was automated, building a dataset of activities was still a long process, as it took significant time to connect into the Appium Server, load up the application on the device and extract the features. We searched the Google play store for relevant applications by using appropriate search terms (e.g., 'Mail', 'Browser', 'To do', etc) and picking the ones with the largest number of downloads. We then download each application into our device, and manually labeled each of the different activities to the activities types we have defined. This manual annotation of activities, was very time consuming, taking roughly about 100 human hours. Most of the activities screens were easily classified to one of the types we have defined, the rest are of different types which we did not model in this paper, and therefore were omitted. This process resulted in a dataset consisting of 80 activities, taken from 50 different applications from the Google play store. 

\subsection{The Classifier}
\label{sec:classifier}
In order to construct a classifier we used Weka \cite{Weka} which is a suite of machine learning software written in Java and is widely used in the machine learning community. In order to train our classifier, we ran a 10-fold classification process on our dataset with different classification algorithms, while measuring the accuracy of each one. Table \ref{tab:accuracy} shows accuracy averaged over activity type prediction using different classification models:
\begin{table}[H]
\begin{center}
 \begin{tabular}{|| c | c ||} 
 \hline
 Classifier & Accuracy  \\ [0.5ex] 
 \hline
 Decision Tree & 63.75\% \\
 \hline
 K-Nearest Neighbours & 77.5\% \\
 \hline
 Logistic Regression & 77.5\% \\
 \hline
 Random Forest & 82.5\%  \\
 \hline
 Multi-Layer Perceptron & 83.75\%  \\
 \hline
 KStar & 86.25\%  \\ 
 \hline
\end{tabular}
\end{center}
\caption{Accuracy of activity type prediction using different classification models.}
    \label{tab:accuracy}   
\end{table}

As we can see from Table \ref{tab:accuracy}, using the instance-based  KStar classifier \cite{cleary1995k} we have managed to achieve a high classification accuracy of 86.25\%, while other classic methods such as decision trees or multi-layer perceptron averaged only 77\% accuracy. Thus, our model of choice for this work is KStar. KStar uses an entropic distance measure as a similarity function to determine which of the training instances are the most similar to the test instance. We performed a grid-search over the possible $k$ parameter values and found that $k=20$ produced the best results.

Additionally, we performed a feature selection process in order to see which features contribute more information to our model. We ran an information gain based feature selection \cite{karegowda2010comparative}, which evaluates the \say{worth} of a feature by measuring the information gain with respect to the class. Table \ref{tab:featuresRanking} shows the ranking of each feature based on its information gain. 

\begin{table}[H]
\begin{center}
 \begin{tabular}{|| c | c ||} 
 \hline
 Feature & Score  \\ [0.5ex] 
 \hline
 number of clickable elements in the middle section of the screen & 1.176 \\
 \hline
 number of general elements & 0.999 \\
 \hline
 number of long-clickable elements & 0.978 \\
 \hline
 number of clickable elements in the bottom section of the screen & 0.802  \\
 \hline
 number of clickable elements in the top section of the screen & 0.78  \\
 \hline
 number of vertical swipeable elements in the middle section of the screen & 0.57  \\ 
 \hline
 is the activity contain a navigation drawer & 0.569  \\ 
 \hline
 number of text fields elements in the middle section of the screen & 0.544  \\ 
 \hline
 number of horizontal swipeable elements in the middle section of the screen & 0.284  \\ 
 \hline
 number of text fields elements in the bottom section of the screen & 0.257  \\ 
 \hline
 number of horizontal swipeable elements in the top section of the screen & 0.234  \\ 
 \hline
 number of horizontal swipeable elements in the bottom section of the screen & 0  \\ 
 \hline
 number of vertical swipeable elements in the bottom section of the screen & 0  \\ 
 \hline
 number of vertical swipeable elements in the top section of the screen & 0  \\
 \hline
 number of text fields elements in the top section of the screen & 0  \\ 
 \hline
\end{tabular}
\end{center}
\caption{Score of each feature based on its information gain (the higher the more informative).}
    \label{tab:featuresRanking}   
\end{table}

Observing the results in Table \ref{tab:featuresRanking}, we can notice that the features related to the number of clickable elements in different sections of the screen, as well as the number of general elements, proved to be very informative in classifying an activity to a type. However, the features related to the swipeable elements in the top and bottom sections of the screen, along with the number of text fields elements in the top section, were not able to contribute any information to the model.

\section{Empirical Evaluation}
\label{sec:evaluation}
\subsection{Experimental Design}
We evaluate our approach against the classic random-testing approach. The Android Application Monkey \cite{Monkey} was chosen as the representative of present tools, considering it being one of the most frequently used tool to test Android applications \cite{choudhary2015automated}. This is attributable to the fact that it is part of the Android developers toolkit. With the purpose of demonstrating the limitations of current tools, as well as showing that our approach of activities classification using machine learning can overcome these limitations, we designed a novel experiment which focuses on applications \textit{logical} bugs. These bugs are related to the application's logic, meaning unwanted behavior in the application's functionality, as opposed to real-time application crashes which are caused by uncaught exceptions thrown in the code.

We used 2 new open source Android applications for the experiment. These applications were not used in this study thus far. Applications usually go through rigorous testing before they are uploaded to the application store. Therefore, in order to simulate a large variety of realistic logical bugs, we artificially \say{planted} bugs in addition to existing ones as discussed next: 

  \begin{itemize}[itemsep=5pt]
  \item \textbf{\say{K-9Mail}} - An email client application. We focused on the following activities:
 \begin{itemize}[itemsep=0pt]
 \item \say{\textbf{MessageList}} - A mail activity which contains the following bugs:
 \begin{itemize}[itemsep=0pt]
\item A user cannot open an email's content from the inbox list.
\item A user can send an email without recipient's address.
\item A user cannot send a valid email.
\item A user can send an email with an invalid recipient's address (this bug already existed in the original code).
\end{itemize}
\item \say{\textbf{setup.AccountSetupBasics}} - A login activity  which contains the following bugs:
\begin{itemize}[itemsep=0pt]
\item A user can sign in without filing in a username and a password.
\item A user can sign in with an invalid username and an invalid password.
\item A user cannot sign in with a valid username and a valid password.
\end{itemize}
\end{itemize}
\item \textbf{\say{CrimeTalk Reader}} - A portal application to browse \say{CrimeTalk} articles. We focused on the following activity:
\begin{itemize}[itemsep=0pt]
\item \say{\textbf{MainActivity}} - A portal activity which contains the following bugs:
\begin{itemize}[itemsep=0pt]
\item A user cannot swipe the screen left and right in order to browse the portal's different sections.
\item A user cannot click on the menu's different tabs in order to browse the portal's different sections.
\item A user cannot open an article from the activity.
\end{itemize}
\end{itemize}
\end{itemize}

 The ACAT testing tool was configured to run for 2 minutes, while the Android Monkey was set to invoke 50,000 pseudo-random user interface events on the activity, which is approximately equivalent to running a test for 2 minutes as well. We ran both conditions on the original activity, which has not been tampered with, and on the faulted version thereof. Finally, we extracted the results of each run's report, which includes the number of crashes discovered, the number of logical bugs discovered and, for our add-on, the classification of each activity as well.

\subsection{Results}
\label{sec:results}

\begin{figure} [H]
  \includegraphics[scale=0.28]{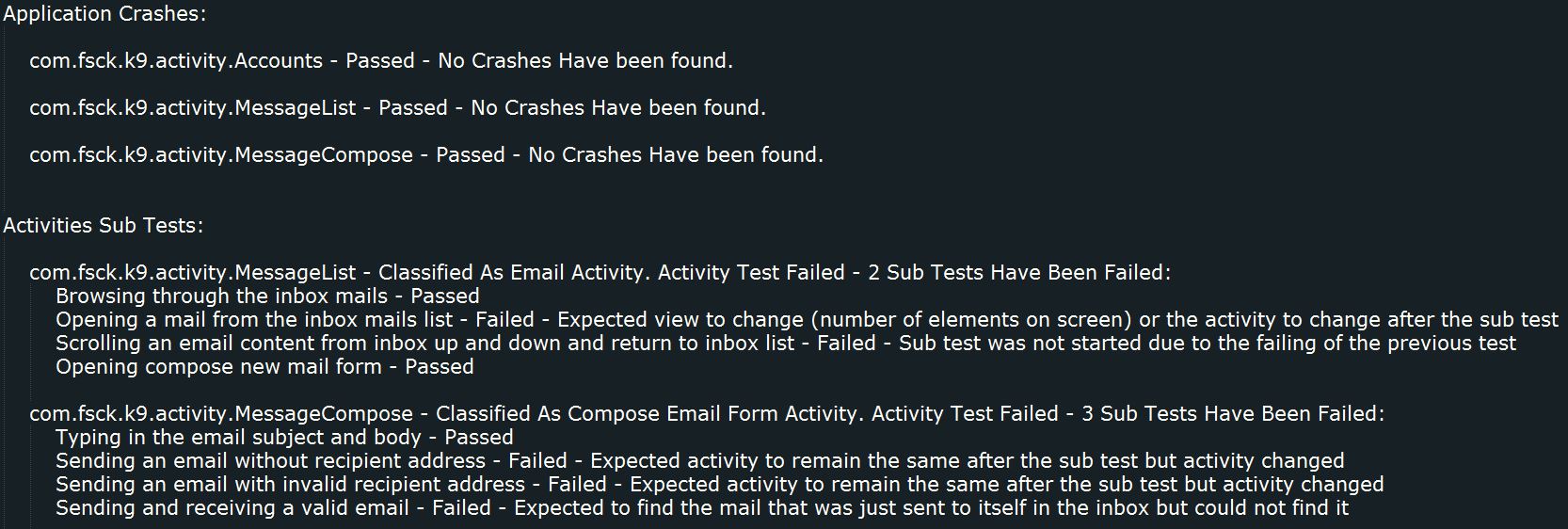}
  \caption{An excerpt from the ACAT’s produced report which describes the discovered bugs.}
  \label{fig:report}
\end{figure}

\begin{figure}
  \includegraphics[width=\linewidth,scale=8]{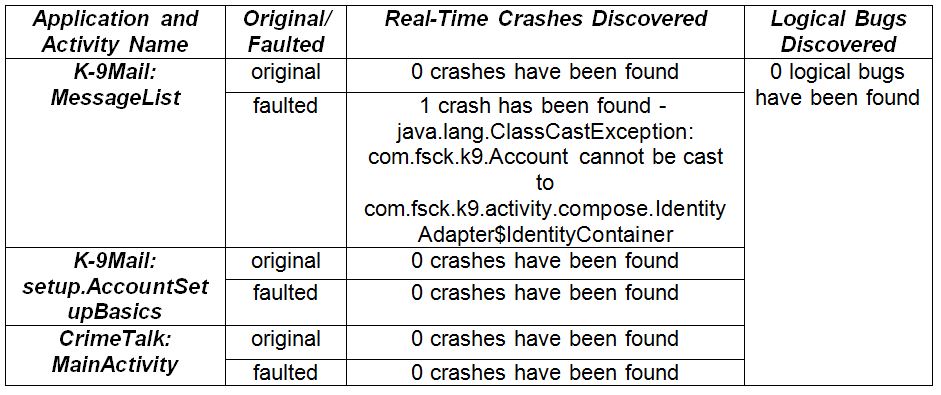}
  \caption{Bugs discovered in the different activities by the Android Monkey tool.}
  \label{fig:monkeyResults}
\end{figure}

\begin{figure}
  \includegraphics[width=\linewidth]{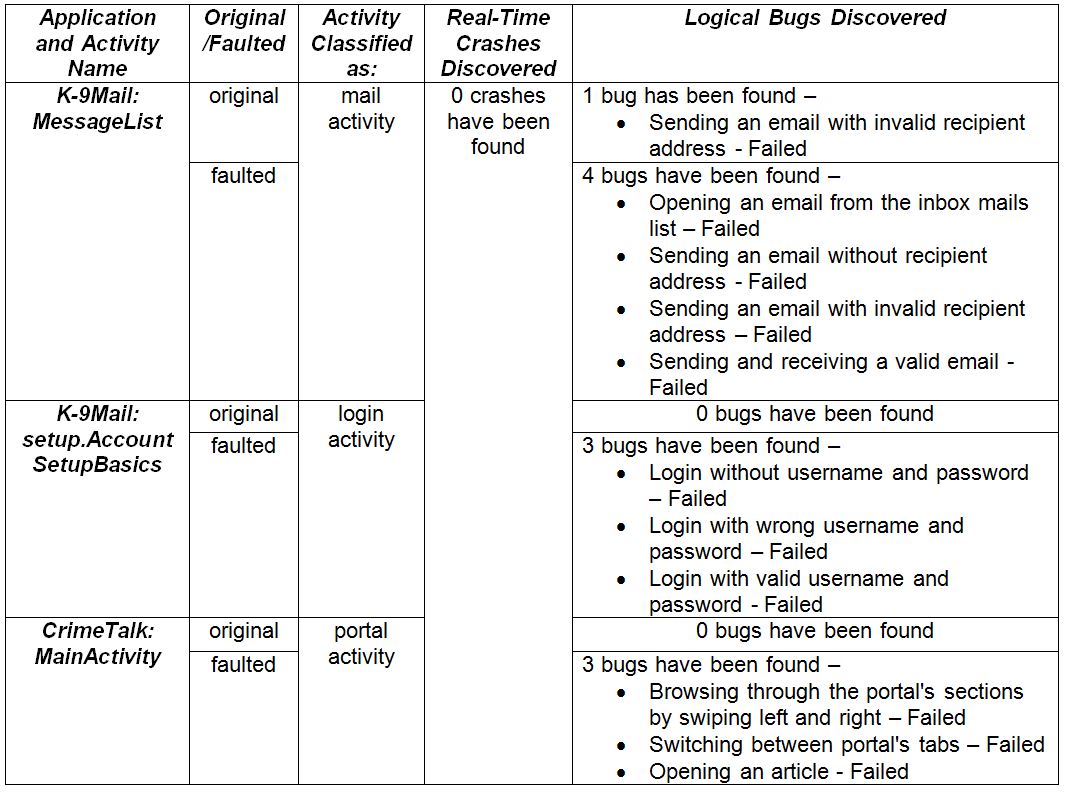}
  \caption{Bugs discovered in the different activities by the ACAT add-on}
  \label{fig:ourToolResults}
\end{figure}

Figures \ref{fig:monkeyResults} and \ref{fig:ourToolResults} present the results of the experiment. While examining them, we can identify 3 major trends: 1) The ACAT was able to classify correctly the 3 unseen activities. 2) The ACAT managed to discover all of the \say{planted} bugs while the Android Monkey discovered none. Moreover, the ACAT was able to find a logical bug (A user can send an email with an invalid recipient's address) which was already part of the original version of the application, without tampering with its code. In the rest of the original activities, it has correctly proclaimed that there are no logical bugs. 3) The Android Monkey managed to discover 1 real-time crash which was not caught by the ACAT. The ACAT also produces a  report which can be seen in Figure \ref{fig:report}.


\section{Discussion}
\label{sec:discussion}
Our results depicted in Section \ref{sec:results} show an interesting phenomenon. While the Android monkey was not able to detect a single logical bug, the ACAT discovered all of the various bugs implemented in the source code of the applications, as well as a bug that already existed in the original code. This is contributed to our activities classification approach, which enables this add-on the power to derive a list of activity-based tests for examining the activity's expected behavior. These tests can only be executed in the correct context, which can be interpreted using this proposed machine learning approach. In addition, our experiment demonstrates another underlying idea behind our approach; instead of testing an application as a whole unit, as done by previous works as described in Section \ref{sec:relatedwork}, it might be better, or at least grant a certain advantage, to consider each activity in the application as a self-entity with specific desired functionality. Thus, an application test could be a series of scenario tests, designed for each activity on its own. To our knowledge, this is the first attempt to develop a machine-learning-based automatic tool for discovering such application bugs.

When presenting a new approach, it is worth to discuss its limitations. Since a system may have an infinite number of possible runs, checking the behavior of an application against our expectations is limited to those executions that we actually carry out. Thus, our approach is limited to the activity types and the functionalities which have been pre-defined. The $7$ activities types identified in this work were developed for a \say{proof of concept} of our approach, the full intended product will contain more types and tests. Additionally, one must consider the preliminary scope of our experiment (only 2 applications) and the fact that we planted bugs in advance. We are currently working with TestProject R\&D team to expand our approach for more activities types, along with enabling the testing algorithm to be less dependent on hard-coded test cases.

\section{Conclusions}
\label{sec:conclusions}
This paper introduces a novel approach for testing Android applications using machine learning techniques. The use of such techniques enabled us to classify each of the application activities into a specific type, which in turn allowed us to test various expected behaviors of the different screens. Furthermore, we have tested our add-on on different applications, demonstrating its advantage against the popular Android applications testing tool -- the Android Monkey. Our add-on, which we named ACAT, is shown to find more logical bugs in an application, as opposed to only real-time crashes, which opens the possibility for developing  more sophisticated testing tools. We are currently working with TestProject in order to integrate the ACAT add-on in the TestProject framework, utilizing their database of thousands of mobile applications patterns. The ACAT add-on will be available to install via TestProject Add-ons store.

\bibliographystyle{ieeetr}

\bibliography{MRM}

\end{document}